 \documentclass[preprint]{aastex}

\newcommand{\etal}{{\em et al.}}

\input{psfig}

\slugcomment{To appear in Astronomical Journal}

\shorttitle{Globular Clusters in groups}
\shortauthors{Da Rocha \etal}

\begin{document}


\title{Globular Clusters around Galaxies in Groups}


\author{Cristiano Da Rocha, Claudia Mendes de Oliveira}
  \affil{Instituto de Astronomia, Geof\'{\i}sica e Ci\^encias Atmosf\'ericas, 
  Universidade de S\~ao Paulo, \\ Av. Miguel Stefano 4200, 04301-904, 
  S\~ao Paulo -- SP, Brazil}
  \email{rocha@iagusp.usp.br, oliveira@iagusp.usp.br}

\author{Michael Bolte}
  \affil{UCO/Lick Observatory, Department of Astronomy and Astrophysics, \\
  University of California, Santa Cruz, California 95064}
  \email{bolte@ucolick.org}

\author{Bodo L. Ziegler}
  \affil{Universit\"atssternwarte G\"ottingen, Geismarlandstr. 11, 37083 
  G\"ottingen, Germany}
  \email{bziegler@uni-sw.gwdg.de}

\and

\author{Thomas H. Puzia}
  \affil{Universit\"atssternwarte M\"unchen, Scheinerstr. 1, D-81679
  M\"unchen, Germany}
  \email{puzia@usm.uni-muenchen.de}

\begin{abstract}
We have obtained deep photometry of NGC 1199 (in the compact group HCG 22)
and NGC 6868 (in the Telescopium loose group) with the Keck II and the
VLT-I telescopes. Both galaxies are the optically brightest galaxy of
their groups. NGC 1199 has two companion galaxies at a median projected
distance of only $33~kpc$ and, based in its peculiar internal structure
and large X-ray halo, NGC 6868 has been proposed to be a merger remnant.

Our analysis of $B$ and $R$ images uncovered a population of globular
clusters around both galaxies, with total (and local) specific frequency
$S_N = 3.6\pm1.8$ ($3.4\pm1.5$) for NGC 1199 and $S_N = 1.8\pm1.1$
($0.8\pm0.4$) for NGC 6868.  The radial profile of the globulars
of NGC 1199 follows the light distribution of the galaxy and can be
fitted by a power--law and a ``core model'' with a very steep slope
($\alpha = 2.5\pm0.3$).  In the case of NGC 6868, the profile of the
globulars is well fitted by a power--law and a ``core model'' profile of
slope $1.4\pm0.3$ and is shallower than the galaxy light distribution.
Maximum-likelihood fitting of two Gaussians to the globular cluster
color distribution yields a high significance for multi-modality with
peaks centered at $(B-R)_0 = 1.13\pm0.04$ and $1.42\pm0.04$ (NGC 1199)
and $(B-R)_0=1.12\pm0.07$ and $1.42\pm0.07$ (NGC 6868).

NGC 1199 and NGC 6868 are good examples of galaxies where the group
environment are likely to have affected their dynamical evolution. We
find that for NGC 1199, the properties of the globular cluster system are
similar to those for other systems around external elliptical galaxies
located in less dense environments, but with a very steep radial profile.
In the case of NGC 6868, we find a regular radial profile and color
distribution and a comparatively low specific frequency for the globular
cluster system of the galaxy.

\end{abstract}



\section{Introduction}

Variations in the properties of extragalactic globular cluster systems
(GCS) as a function of the environment of the host galaxy hold some
of the keys to understanding the formation and evolution of GCS.
Environment has been proposed as one of the important factors in setting
the specific frequency $S_N$ (number of clusters normalized by galaxy
luminosity). \citeauthor{har91}, in a review paper of \citeyear{har91},
pointed out that elliptical galaxies in small groups and sparse
environments, have, on average, half the $S_N$ of elliptical galaxies
in the Virgo and Fornax clusters (excluding the central galaxies which
have the well known extremely high specific frequencies).  On the
other hand \citet{djo92} found an increase of the $S_N$ with the host
galaxy luminosity, what was confirmed by \citet*{zep94} \citep[see][for
detailed discussion on the subject]{ash98,elm00,har01}.  As the majority
of the very bright galaxies are found in clusters and the faint ones
are in groups and in the field, the environmental effect on the $S_N$
and its luminosity dependency are not easily disentangled.  The effects
of host-galaxy environment on other details of the GCS, the radial and
color distributions is not well documented.

The main goal of this work is to characterize the GCS of two galaxies
in small groups: NGC 1199, the central elliptical galaxy in the compact
group HCG 22 and NGC 6868, a suspected merger remnant in the center of
the Telescopium group. In particular, we are looking for suggestions
that the basic properties of the globular cluster systems around these
galaxies have been modified by the dense environment of the compact
group or the merger event respectively.

The compact group HCG 22 contains three bright galaxies.  This group was
originally cataloged as a quintet of galaxies \citep{hic82}.  However,
\citet{hic92} showed this to be a triplet at a mean redshift of 0.009 with
a superimposed discordant pair at $z = 0.0296$.  The group has a median
velocity of 2686 ${\rm km~s}^{-1}$ and a velocity dispersion of $\sim$
54 $km~s^{-1}$. Based on the surface brightness fluctuation method,
\citet{ton01} derive a distance modulus $(m-M)_V = 32.6\pm0.3$ to NGC
1199, the dominant galaxy of the group, corresponding to $33.1~Mpc$
and a median projected galaxy separation in the group of $33~kpc$
\citep[][assume $H_0=74~km~s^{-1}~Mpc^{-1}$]{ton01}.  NGC 1199 (also
known as HCG 22A) is classified as E2 \citep{hic82}.

NGC 6868 is also classified as an E2 \citep{dev91}. The Telescopium group
contains five bright galaxies with a median velocity of 2601 $km~s^{-1}$,
and a velocity dispersion of $\sim$ 219 $km~s^{-1}$.  \citet{ton01}
measure $(m-M)_V = 32.1\pm0.2$ based on the surface brightness fluctuation
method which corresponds to a distance of $26.8~Mpc$ and a median
projected separation for group galaxies of $209~kpc$.  Although NGC 6868
is located in an environment that has a number density of bright galaxies
lower than that of a typical compact group, the interest in studying the
globular cluster system of this galaxy is that it is almost certainly
a recent merger \citep*{han91}.  NGC 6868 appears to be an ordinary
elliptical galaxy from its global properties, but, in a detailed study,
\citet{han91} showed that there is a system of dust features close to
the NGC 6868 core and a disk of ionized gas within $1~kpc$ of its center
indicating the presence of a young population of stars. The kinematics
of the gas disk was studied by \citet{pla98}, who found a flat rotation
curve with a velocity amplitude of $\pm150~km~s^{-1}$.  The presence of
the dust structure and the ionized gas suggest that NGC 6868 may have
captured one or more gas--rich galaxies.  In addition NGC 6868 has a halo
of X--ray gas associated with it. Its X--ray and optical luminosities are
in good agreement with the $L_B-L_X$ relation proposed by \citet{beu99}
for early--type galaxies. The general properties of the two galaxies
can be seen in table~\ref{tabdata}.

The paper is organized as follows. Section 2 describes the observations
and data reduction. In \S3 we show the results and \S4 has our discussion.

%Table 1
\placetable{tabdata}

\section{Observations and Data Reduction}

\subsection{Observations}

The images of HCG 22 were obtained with the Keck II telescope
in June 1997 using the Low--Resolution--Imaging Spectrometer
\citep[LRIS,][]{oke95}. Images in $B$ and $R$ of total exposure times of
720 seconds (4 $\times$ 180) and 630 seconds (7 $\times$ 90) were obtained
with an average seeing, on the combined images, of 0.77 and 0.74 arcsec
respectively. The field size is $5.7 \times 7.3$ arcmin and the pixel size
is 0.215 arcsec. The NGC 6868 images were obtained with the ESO/VLT-I
in October 1999 using the Focal Reducer/Low Dispersion Spectrograph 1
(FORS1).  Images in $B$ and $R$ with total exposure times of 900 seconds
(5 $\times$ 180) and 810 seconds  (9 $\times$ 90) were obtained with
average seeing values, on the combined images, of 0.76 and 0.73 arcsec
respectively. The field size is $6.4 \times 6.4$ arcmin and the pixel
size is 0.2 arcsec. Images in $B$ and $R$ of similar depth to the ones
taken for NGC 6868 were obtained centered on a position 10$^\prime$ from
the galaxy center.  This was meant to be our control field for background
subtraction.  The log of observations is shown in Table~\ref{tabimag}.

% Table 2
\placetable{tabimag}

\subsection{Data Reduction}

\subsubsection{Image Processing and Photometry}

Using the IRAF\footnotemark~package (Image Reduction and Analysis
Facility) we carried out the basic image reduction procedures (bias
correction, flatfielding and image combining).  We next used the tasks
ELLIPSE and BMODEL of the STSDAS package to model and subtract the light
from bright galaxies. We masked the very bright objects on the images
and the central regions of the giant galaxies that could not be modeled
by ELLIPSE.  Subtraction of the sky level of the image was done using the
SExtractor package (Source Extractor) Version 2.1.6 \citep{brt96} with the
output option --BACKGROUND and a meshsize of 64 pix.  The combined images
for NGC 1199 and NGC 6868 and the final masked and background--subtracted
images are shown in figure~\ref{figima}.

\footnotetext{IRAF is distributed by the National Optical Astronomy
Observatories, which is operated by the Association of Universities
for Research in Astronomy, Inc., under cooperative agreement with the
National Science Foundation.}

% Figure 1
\placefigure{figima}

The combined, masked and background--subtracted images were then used
for detection and photometry of the faint objects, using SExtractor,
with a Gaussian convolution mask of 3 pixels for the HCG 22 images and
of 4 pixels for the NGC 6868 images.  For photometry of the objects
we used the MAG\_AUTO value, which is given by an adaptive aperture
photometry routine, and gives the total magnitude of the detected objects.
The adaptive aperture photometry routine is based on the \citet{kro80}'s
``first moment'' algorithm \citep[for more details on the photometry
see][]{brt96}.  Monte Carlo simulations were made to test the reliability
of the MAG\_AUTO photometry (see following section) and a small correction
was added to the photometry of each detected object.

The final list of objects includes only the objects that were detected in
both $B$ and $R$ bands.  We discarded all the objects that were within 5
arcsec of the borders of the images and within 5 arcsec of masked regions.

At the distance of our groups, globular clusters are point--like
sources.  To select them, we have excluded the extended objects (dwarf
and background galaxies) from our final list. Using the results from the
add--star experiments (described in the next section) we could define the
range of FWHM where we expect the point--like sources to be found, as a
function of their magnitudes. The FWHM value is measured by SExtractor.
Taking the average of the FWHM on the $R$ band, in 0.5 magnitude bins,
of the simulated objects and using a 1 $\sigma$ rejection level, we could
determine a ``point--like source locus'' on the magnitude--FWHM plane
for each of the images. The detected objects located inside this region,
along the magnitude range for each image, were classified as point--like
sources and all the other objects were excluded from the final list. This
method has shown, in our case, better results than the default SExtractor
star--galaxy separations. The selection of point--like sources using
the magnitude--FWHM plane was previously used to select GC candidates
by \citet{rho01}. The magnitude--FWHM plots for our fields are shown
in figure~\ref{figfwhm}.

% Figure 2
\placefigure{figfwhm}

Calibration of the images of HCG 22 was done using images from the CTIO
1.5--meter telescope. We have calibrated those HCG 22 images in $B$ and
$R$ bands using \citet{lan92} standard stars and then calibrated the Keck
II images using the surface brightness profile of NGC 1199. The final
RMS of these calibrations are 0.011 for the $B$ band and 0.016 for the
$R$ band. We have checked the calibration using the surface brightness
profiles of NGC 1199 shown in \citet{men92} and no zero point differences
were found. For NGC 6868 a similar procedure was applied using images
from the Las Campanas Observatory 2.5--meter telescope and calibrated
with \citet{lan92} standard stars.  The RMS of these calibrations are
0.003 for the $B$ band and 0.004 for the $R$ band. In order to obtain
the standard star photometry, we used the packages APPHOT and DAOPHOT.

\subsubsection{Add--star experiments}

We used each final image, with bright objects masked out, for an add--star
experiment using the IRAF task ADDSTAR in the DAOPHOT package.  Our goals
were to evaluate the detection completeness, photometric errors and the
limit of the Star--Galaxy (SG) separation method.

We added 600 artificial stellar--like objects to each 0.5 mag bin in
the $B$ image, covering the range $18<B< 28$.  In the $R$ image we
have added objects in the same position of the ones added to the $B$
image, with $(B-R)$ colors ranging randomly from 0 to 2.5.  The objects
were generated from the PSF for each field. The PSF was constructed
using typically 40 to 80 point--sources found in the images. These
artificial point sources were added in 30 runs of 20 objects each, to
avoid artificial crowding of the field. Detection and photometry of the
objects in the frames which contained the artificial objects was then
performed in the same manner done for the original images. The final
simulated object list is composed by the simulated objects recovered in
both bands. The number of objects which were recovered over the input
number then gave us the completeness fraction.  A comparison of the
input and output magnitudes of the objects gave us estimates of the
random and systematic uncertainties in the photometry.

We have determined the completeness fraction in radial rings centered
on the galactic center, and this radial completeness fraction was the
one applied in our analyses. This radial dependence is expected since
the noise increases in the galaxy subtracted image as we get close to
the center of the galaxy. The radial completeness fractions for both
galaxies can be seen in figure~\ref{figcompl}.  We have also studied
the completeness as a function of color and found the variations being
at most 5\%.

% Figure 3
\placefigure{figcompl}

As can be seen in figure~\ref{figcompl}, the SExtractor magnitudes are
systematically too faint by $\sim 0.05$ mag.  This difference is expected
and is due to the aperture correction applied by SExtractor being slightly
underestimated for point sources \citep{brt96}. We applied a correction
of this light loss to the photometry of detected objects in each of
the frames.  The completeness limit for our data given by the add--star
experiment was considered to be the magnitude at which at least 50\%
of the objects were recovered and is shown in table~\ref{tablim}.

% Table 3
\placetable{tablim}

\section{Analysis and Results}

\subsection{Globular Cluster Luminosity Function}

In order to select the globular cluster candidates, we have restricted
our samples to objects with $22.0<B<25.5$ and $20.0<R<24.0$ for NGC 1199
and $22.0<B<24.5$ and $20.0<R<23.0$ for NGC 6868, which represent the
ranges of magnitudes where we expect to observe GCs (the faint limit is
given by the completeness limit).  We have also applied color limits
$0.7<(B-R)_0<2.1$ which is the range where we expect to find the GCs
\citep{tho93,puz99,woo00}, eliminating very blue and red unresolved
objects.

For our selected sample, we plot the number--density as a function of
distance from the center of the galaxies (see figure~\ref{figrad}). In
those plots we can clearly see at which radius the background level
for the objects was reached.  Counting objects in rings of 0.2 arcmin,
and correcting the counts by incompleteness, we can see that the radial
profile reaches the background counts at distances larger than 2.4
arcmin from the center of NGC 1199. For NGC 6868 we have counted objects
in rings of 0.5 arcmin. The counts in the background field (for radii
greater than 10 arcmin in figure~\ref{figrad}) are approximately constant
and these are taken to represent the background counts for this group.
The errorbars represent the poissonian error on the number counts.

We have used the flat region between 2$^\prime$.4 ($23~kpc$) and
4$^\prime$.2 ($40~kpc$) around NGC 1199 to estimate the background level.
We do not have control fields taken away from the group center to make a
background estimation in another way.  From figure~\ref{figrad} we note
that there is a central concentration of objects within a radius of 2.4
arcmin from the center of NGC 1199. We identify this concentration with
the GCS of this galaxy. A similar analysis was applied to NGC 1190 (HCG
22B) and NGC 1189 (HCG 22C) and no significant central concentration
of point--like sources was found around them. From their morphological
types (Sa and SBcd) and their luminosities ($M_V=-18.7$ and $M_V=-19.6$,
\citealt{hic89}) they are not expected to have a large GC population
\citep{har91,ash98}.

% Figure 4
\placefigure{figrad}

We can also see a high concentration of objects around the center of NGC
6868, as is shown in figure~\ref{figrad}. At galactocentric distances
larger than 10 arcmin, $78~kpc$, (objects located on the second field),
we have a flat region which is used as the background control field.
We regard as our sample of possible candidate globular clusters of
NGC 6868 the objects within a distance of 3.8 arcmin ($30~kpc$) of the
center of the galaxy (within the magnitudes and color limits previously
defined). The objects within an annulus of inner radius of 10 arcmin
and outer radius 17 arcmin were considered to belong to a sample of
background galaxies and foreground stars (within the same limits).

We will, hereafter, refer to the area within 2.4 arcmin from the center of
NGC 1199 and 3.8 arcmin from the center of NGC 6868 as the ``on--galaxy
region'' or ``studied area'' and to the region within 2.4 and 4.2 arcmin
from the center of NGC 1199 and the field taken 10 arcmin from the center
of NGC 6868 as the ``background area''.

Since we do not have an independent background field for NGC 1199,
we have estimated the number of foreground stars in the direction of
the galaxy using the galactic model by \citet*{san96}.  We found that
in the magnitude range of the globulars, the possible contamination of
our sample by foreground stars is $0.4~{\rm objects/arcmin^2}$, which
gives us only 5.5 objects in the studied area.  This foreground stars
estimate is negligible and the number of GCs found around NGC 1199
could not be much affected by a wrong stellar contamination estimate
in our small area background field. However, by far the largest source
of contamination is by background galaxies.  The background field used
for NGC 1199 (the outskirts of the field) gives a statistically more
uncertain background subtraction and can still contain bona--fide GCs
that will be counted as background objects (although figure~\ref{figrad}
shows a very flat profile beyond a distance of 2.4 arcmin from the center
of the galaxy).  In the case of NGC 6868, since the estimate of the number
of foreground/background objects was done using an independent frame of
same size as the on--galaxy frame, obtained at a distance of 10 arcmin
to the center of the galaxy, the background subtraction is more reliable.

The globular cluster luminosity function (GCLF) was built binning the
GC candidates to 0.5 magnitude. The number counts were corrected for
incompleteness dividing the number counts of each bin by the completeness
fraction.  The photometric errors were neglected and the errors for the
corrected number counts were given by

\begin{equation} 
\sigma^2 \approx [\frac{N_{obs}}{f^2} + \frac{(1-f)N^2_{obs}}{N_{add}f^3}] 
\end{equation} 
\citep{bol94}, where $N_{obs}$ is the number of objects detected in
the image, $N_{add}$ is the number of objects added to the image in
the addstar--experiment and $f$ is the completeness fraction to this
magnitude bin.

The corrected number counts for the on--galaxy region were subtracted by
the corrected background number counts normalized to the sampled area.
For NGC 1199, we have selected $128\pm11$ objects, which corrected for
incompleteness gives $158\pm14$ objects, in an area of 13.7 arcmin$^2$. In
the background, for the same area, we have a corrected number count
of $61\pm10$ objects, leaving a net number count of $97\pm18$ objects.
For NGC 6868, there are $111\pm10$ selected objects, which corresponds
to $124\pm12$ objects when corrected for incompleteness, in an area of
27.2 arcmin$^2$.  In the background we have a corrected number count
of $74\pm8$ objects in the same area resulting in a net number count of
$50\pm14$ objects.

The luminosity function of globular clusters around giant elliptical
galaxies has been well studied in the literature.  It is generally
accepted that it can be well approximated by a Gaussian distribution
function.  For the peak (turnover magnitude -- $m_0$) and the dispersion
we have used the typical values for elliptical galaxies $M_V =
-7.33\pm0.04$ and $\sigma = 1.40\pm0.05$ \citep{har01}.

The galactic extinction is the same for both galaxies, $E(B-V) = 0.056$
\citep*{sch98}. With the reddening laws found in \citet{rie85} we have
$A_B=0.23$, $A_R=0.13$ and $E(B-R) = 0.10$.

Using the distance moduli for our galaxies and the reddening
corrections above, we expect to start seeing the brightest GCs for NGC
1199 at $B=22.5$ and $R=20.9$ and the turnover at $B=26.2\pm0.3$ and
$R=24.6\pm0.3$.  For NGC 6868, $B=22.1$ and $R=20.5$ for the brightest
GCs and $B=25.7\pm0.2$ and $R=24.1\pm0.2$ for the turnover magnitude.
As we do not reach the turnover point of the GCLF with our data, we did
not attempt to fit the peak and the dispersion values, leaving only
the normalization of the Gaussian as a free parameter. The $\chi^2$
of the fits and the area of the Gaussians covered by our data for both
galaxies are seen in table~\ref{tabgauss} and figure~\ref{figlf}.

% Figure 5
\placefigure{figlf}

% Table 4
\placetable{tabgauss}

\subsection{Specific Frequency and Radial Profile Modeling}

The specific frequency ($S_N$), as defined by \citet{har81}, is the
number of objects normalized by the galaxy luminosity.

\begin{equation} 
S_N \equiv N_{GC} 10^{0.4(M_V+15)} 
\end{equation}

Using a Gaussian fit for the GCLF we can determine the part of the
GC population that we are measuring and extrapolate the number of GCs
over all magnitudes in our image. With the number of GCs found in the
on--galaxy areas (local number) and the light in the same area where
the GCs were counted, we can calculate the ``local specific frequency''
($S_{N_l}$), $i.e.$, the specific frequency of the GCs that are detected
in our images.

Correcting the number counts of NGC 1199 for the unobserved part of the
GCLF we have a total number of $314\pm105$ GCs over all magnitudes.
The galaxy light, estimated using the ELLIPSE model in the same
area, is $V=12.9\pm0.2$ or $M_V=-19.9\pm0.3$.  Using those values to
calculate the local specific frequency ($S_{N_l}$) for NGC 1199 we have
$3.4\pm1.5$. Applying the same procedure for NGC 6868 the local number
of GCs over all magnitudes is $266\pm106$ and the light in this area is
$V=11.0\pm0.2$ or $M_V=-21.3\pm0.3$, which gives a $S_{N_l}=0.8\pm0.4$.
The errors on $S_{N}$ include the error on the number counts, background
subtraction, GCLF extrapolation and its parameters uncertainties, the
photometric error on the galaxy magnitude and the error in the distance
of the galaxy.

We now determine the global specific frequency ($S_{N_g}$) for the system.
$S_{N_g}$ is calculated using the extrapolation of the radial profile
to estimate a total number of GCs for each galaxy and the total light
of the galaxy.

To estimate the total number of objects in the GCS, the sample was
divided in radial rings of 0.2 arcmin for NGC 1199 and 0.5 arcmin
for NGC 6868. The number of GC candidates in each radial ring was
corrected by the lost area due to the masks and unobserved regions. We
then calculated the total number of GCs over all magnitudes in the
complete ring from the central mask and the limiting galactocentric
distance. We use two different radial profiles.  The first one is
a variation of the ``core model'' profile proposed by \citet{for96}
($\rho=\rho_0(r_c^{\alpha} + r^{\alpha})^{-1}$) with the core radius
estimated for the GCSs of elliptical galaxies by the relation shown in the
same work ($r_c(kpc)=-(0.62\pm0.1)\cdot M_V-11.0$). This ``core model''
profile is a simplified analytic form for the King profile \citep{kin62}.
The other profile is a power--law profile ($\rho \propto r^{-\alpha}$).

For either profile the best--fitting model was determined and the total
number of GC estimated by integrating the profiles.  For NGC 1199 the
profile was constrained by the corrected number counts between 0.45
and 2.4 arcminutes. For NGC 6868 the fitted range was 0.35 to 3.8
arcminutes. Table~\ref{tabsn} and figure~\ref{figking} summarize the fits.

For NGC 1199 the slope of the power--law profile is $2.5\pm0.3$, which
is very steep and therefore the total number of GCs is dominated by
the central region. To be able to integrate this profile, we have set a
minimum inner radius of 0.1 arcmin ($\sim~1~kpc$). We assume that there
is no contribution inside this radius since the GCs in this region were
probably destroyed by erosion processes. This mathematical artifact gives
an upper limit of the total population of GCs.  For NGC 6868 the slope of
the best--fit power--law profile is $1.4\pm0.3$ and it can be integrated
analytically (see results below). The slopes used in the ``core model''
profile are the same used in the power--law profile.

The total number of GCs for each galaxy estimated by the two kinds of
profiles can be seen in table~\ref{tabsn}, and the radial profiles of
GC candidates with the power--law and ``core model'' profiles used
and the galaxy light overplotted are shown in figure~\ref{figking}.
The assumption on the profile shape is very important because we have to
extrapolate the number counts in the outer and inner parts of the galaxy.

The profiles were extrapolated to $100~kpc$, which is the maximum radius
we expect the GCS to extend. \citet{rho01} do not detect GCs farther
than $100~kpc$ in NGC 4472, which is at least twice as luminous as
the galaxies in our sample. Three sources of error were considered on
the radial profile extrapolation using the ``core model'': the errors
on the slope, on the core radius and on the outer radius cut--off. To
estimate the error we make by cutting the profiles at the outer radius
($100~kpc$) we have calculated the variation of the results by varying
the outer radius cut--off by 20\%.  For NGC 1199, where the numbers are
dominated by the inner region, the error on the $S_N$ is 0.06 and for NGC
6868 is of 0.2. The error introduced in the $S_N$ by a 30\% variation
of the core radius is of 0.6 for NGC 1199 and only 0.04 for NGC 6868,
where the numbers are dominated by the outer region.

For the extrapolation using the power--law profile three sources of error
were also considered: the error on the slope and on the outer radius
cut--off, in the same manner as for the ``core radius'' extrapolation,
and the error on the inner radius cut--off, for the case of NGC 1199. We
have varied the inner radius cut--off ($1~kpc$) by 50\% to estimate the
error we make on the cut, which translated into an error for the $S_N$
of 1.8. The errors caused by all the sources, for both models, were
added in quadrature and are shown in table~\ref{tabsn}.

% Figure 6
\placefigure{figking}

% Table 5
\placetable{tabsn}

To calculate the $S_{N_g}$ we have used the total magnitude of the
galaxies and the total number of GCs obtained through the integration
of the two kinds of profiles. The results are shown in table~\ref{tabsn}
and are discussed in section 4.

\subsection{GC Color Distribution}

Bimodal color distributions of GCs are found in at least 50\% of
elliptical galaxies \citep{geb99,kun01} and even in some spiral galaxies
like our own and M31 \citep[][and references there in]{ash98,har01}.
They are thought to be due to multiple events of star formation in the
galaxies' histories.  This effect was predicted in elliptical galaxies
by \citet{ash92} merger model, first detected in NGC 4472 and NGC
5128 \citep{zep93b} and since then other models appeared to explain
it, like the multiple collapse model \citep*{for97} and the accretion
model \citep*{cot98,hil99}. Since we are searching for modifications on
the GCS properties caused by the small group environment, multimodal
color distributions of GCs may give some useful information about the
GCS history.

The color distribution of all the objects detected in the on--galaxy areas
is analyzed here, with no attempt to perform a background subtraction. The
color distributions of the background objects, which can be seen in
figure~\ref{figcolorbkg}, are very broad with one dominant peak. The peak
positions are not located at the same place as the GCs candidates peaks
(see figure~\ref{figcolor}). Therefore, we do not expect the shape of
the background color distribution to affect our analysis.

We have applied a KMM test \citep*{ash94} to detect and estimate
the parameters of possible bimodalities in the color distributions.
The KMM code uses maximum likelihood to estimate the parameters that
best describe the sample distributions for a single Gaussian and for
multiple Gaussian fits and calculates the probability of the single
Gaussian being the best fit for the sample distribution. This code
can be run in two modes, homoscedastic, which finds groups with the
same covariance values, and heteroscedastic, which finds groups with
dissimilar covariance values. The last mode has reliability problems
with the analytic approximation to the significance level and for this
reason it is not commonly used for this kind of analysis.

Running KMM on our color distributions, with the homoscedastic case,
yields the two--Gaussian option as the result with 99.9\% confidence
level for both galaxies.  For NGC 1199 the peaks are located at
$(B-R)_0=1.13\pm0.04$ and $(B-R)_0=1.42\pm0.04$ with 62\% of the GCs on
the blue peak and 38\% on the red one and a covariance of 0.01. For NGC
6868 the values found are $(B-R)_0=1.12\pm0.07$ and $(B-R)_0=1.42\pm0.07$
with 51\% of the GCs on the blue peak and 49\% on the red peak and a
covariance of 0.009. The errors represent the mean error on the peak
position given by the KMM added in quadrature with the mean photometric
errors of the colors.

A visual inspection of the color histograms and the estimates using the
Epanechnikov kernel density estimator \citep[see][for detailed discussion
on the kernel density estimator]{sil86} shows that there might be a
bimodality for NGC 1199 and that there are, at least, two major peaks for
NGC 6868. The KMM results, bimodality significance and the peak positions,
are in good agreement with the visual inspections and the kernel density
estimates, as can be seen in the distributions in figure~\ref{figcolor}.

Splitting the GC candidates into blue and red sub--samples, we have
analyzed the radial distribution of each sub--population and found that
the red clusters are more centrally concentrated than the blue ones,
as found in the literature \citep*{gei96,kis97,lee98,kun98}. The slopes
found fitting a power--law to the radial profiles are $2.2\pm0.4$
and $2.9\pm0.6$ to the blue and red sub--populations of NGC 1199,
respectively, and $0.8\pm0.4$ and $1.7\pm0.2$ to the blue and red
sub--populations of NGC 6868, respectively.

% Figure 7
\placefigure{figcolorbkg}

% Figure 8
\placefigure{figcolor}

\section{Discussion and Conclusions}

We present a short summary of the results we have found so far before
we proceed with the discussion:

(1) We detect a significant population of centrally concentrated
GCs around the galaxies NGC 1199 and NGC 6868, as shown in
figure~\ref{figrad}.

(2) The two first--ranked group galaxies studied here show different $S_N$
values. For NGC 1199, the $S_N$ appears to be ``normal'' for an elliptical
galaxy, while for the suspected merger NGC 6868, the specific frequency
is somewhat low compared to other galaxies of similar luminosities.

(3) We find a good case for bimodal color distributions in the populations
of GCs and in both cases the radial distributions for the red objects
have steeper slopes than those for the blue objects.

In the following we will further discuss these main results and will
attempt to fit our observations into what is known about GC systems
around other elliptical galaxies.

\subsection{Radial profiles and population sizes}

The radial profiles shown in figure~\ref{figrad} show a central
concentration of objects within 2.4 arcmin ($23~kpc$) of the center of
NGC 1199 and 3.8 arcmin ($30~kpc$) of the center of NGC 6868, indicating
the existence of a significant population of globular clusters around
these galaxies.

NGC 1199 has a very steep radial profile that follows the galaxy light
and can be fitted by a power--law with $\alpha=2.5\pm0.3$ and by a ``core
model'' profile with the same slope and a core radius of $r_c=2.19~kpc$.
The light profile of NGC 1199 is as steep as the radial distribution of
GCs out to a radius of 1.5 arcmin and then it becomes steeper.  Using our
determination of the global specific frequency for this galaxy done with
the ``core model'' profile, our value of S$_N$ can only be taken as an
upper limit, if the real profile of the GCSs follows the light of the
galaxy exactly.  The steep profile can be caused by the destruction of
the outer part of the system by tidal effects, since we have two giant
galaxies closer than $35~kpc$ and with small relative velocities,
or by some inefficiency of the erosion processes in the inner part
of the galaxy.  The global specific frequency estimated ($3.6\pm1.8$
using the ``core model'' and $5.2\pm3.2$ using the power--law model)
is similar to the ``normal'' value for elliptical galaxies, where the
``normal'' value is defined as a typical value for elliptical galaxies
as given by \citet{har01}.  There is no optical sign of star formation
regions in the galaxy that could artificially change the $S_N$ value.

The GCS of NGC 6868 has a regular radial profile, the slope of the
power--law fitted to the distribution is $\alpha=1.4\pm0.3$. The slope
value for a power--law radial profile of GCs for a normal elliptical
galaxy ranges from 1 to 2 \citep{ash98}.  The GCS profile is more extended
than the galaxy light, which has a slope of 1.72, as can be seen in
figure~\ref{figking}.  Our final result for the global specific frequency
for this galaxy is $S_N=1.8\pm1.1$, assuming a ``core model'' profile,
and $1.9\pm1.0$, assuming the power--law profile.  This value is lower,
by almost a factor of two, than the ``normal'' value of \citet{har01}
for an elliptical galaxy ($S_N \sim 3.5$) and indicates a poor GCS
(although it's well known that the scatter on the values of $S_N$ for
elliptical galaxies is large).  This might be due to a high efficiency
of forming stars in the past as opposed to the formation of globular
clusters or to very efficient destruction mechanisms that could perhaps
be related to the recent merger that may have occurred in the central
part of the galaxy \citep{han91}. As for NGC 1199, there are no obvious
signs of wide--spread star formation throughout the galaxy  that could
affect the $S_N$ value of NGC 6868.

\subsection{Colors of the GC population}

According to stellar population evolutionary synthesis models
\citep{wor94,bru93} metallicity effects are supposed to dominate color
differences in globular clusters if the populations are older than 1
Gyr. We expect the bulk of the populations of GCs to be older than 1 Gyr,
since we could not find evidences of recent star formation.  A linear
relation between the color and the metallicity is predicted by the
population synthesis models as the relation proposed by \citet*{ree94}
($[Fe/H]=3.112(B-R)_0-4.967$~~~$\sigma[Fe/H]=0.21$).

Some studies have found doubly--peaked color distributions of GCSs
around elliptical galaxies. \citet{geb99} and \citet{kun01} found
for samples of respectively 43 and 29 early--type galaxies observed
with HST, that at least 50\% of the elliptical galaxies have GCSs
with bimodal color distributions.  Studies of the GCS of NGC 4472
(M49) \citep{gei96,lee98,puz99} have found color distributions with
peak metallicities corresponding to $[Fe/H]=-1.5\pm0.05~dex$ and
$[Fe/H]=-0.32\pm0.05~dex$, using the color--metallicity relation given
by \citet{kun98} ($[Fe/H]=-(4.50\pm0.30)+(3.27\pm0.32)\cdot(V-I)$).
\citet{nei99}, in a study of 12 Virgo elliptical galaxies found a bimodal
color distribution for the GCSs around 8 of them.

The recent work of \citet{lar01}, presents an analysis of 17 nearby
early--type galaxies, homogeneously observed with HST. In this work
they have found a correlation between the GCS mean color and the
luminosity and the mean color and the velocity dispersion of the host
galaxy. Strong correlations were found for the red peaks, which were also
found by \citet{for97} and by \citet{for01}. The latter study suggested
the following color--velocity dispersion relation $(V-I)=0.23 \cdot
log\sigma+0.61$. Weak correlations were found for the blue peaks
and were discarded by the previously cited works. \citet*{bur01}
have found an average metallicity of $[Fe/H]= -1.4~dex$ for the blue
peak, which agrees with the average value found previously by other
authors. \citet{kun01} found that the correlations for both peaks are,
at best, weak.  In general, the average metallicity of the globular
clusters on a bimodal--distribution, when considered as a single
population, is in agreement with the value of a unimodal distribution
($[Fe/H]=-1.0\pm0.05~dex$, \citealt{kun01}).

The GCSs around the two galaxies studied here were analyzed with the
the KMM code \citep{ash94} which has detected bimodal populations for
the GCSs around NGC 1199 and NGC 6868.

For NGC 1199, KMM has located peaks at $(B-R)_0=1.13\pm0.04$ and
$(B-R)_0=1.42\pm0.04$, which corresponds to $[Fe/H] = -1.45\pm0.21~dex$
and $[Fe/H] = -0.55\pm0.21~dex$, respectively \citep{ree94}, similar
to the values were found for many published galaxies in the last years,
as seen in the following.  Transforming our $(B-R)$ values to $(V-I)$,
using the relation given by \citet{for01} ($(V-I)=0.68(B-R)+0.15$),
and overplotting them (large filled symbols) on the \citet{for01}
color--velocity dispersion plot (figure~\ref{figforbes}), we can see
that our results are consistent with theirs and those of other studies.

For NGC 6868, peaks found by KMM are located at $(B-R)_0=1.12\pm0.07$ and
$(B-R)_0=1.42\pm0.07$, corresponding to $[Fe/H] = -1.48\pm0.22~dex$ and
$[Fe/H] = -0.55\pm0.22~dex$, respectively \citep{ree94}.  Those values are
also in agreement with the ones found for many galaxies in the literature.
In figure~\ref{figforbes} we can see our measured values overplotted
onto the diagram of \citet{for01} (large open symbols).

% Figure 9
\placefigure{figforbes}

The results reported in this work show that there is a need of more work
on globular cluster systems around small--group elliptical galaxies in
order to understand the effects to which the GCSs are exposed in such
environment.

\acknowledgments

CDR is supported by FAPESP (Funda\c c\~ao de Amparo a Pesquisa do
Estado de S\~ao Paulo) PhD. grant No. 96/08986--5. CMO acknowledges
support from FAPESP (Funda\c c\~ao de Amparo a Pesquisa do Estado de
S\~ao Paulo). MB is happy to acknowledge support from National Science
Foundation grant AST 99-01256.  BLZ acknowledges support from the Deutsche
Forschungsgemeinschaft (DFG) and the VW foundation. THP gratefully
acknowledges the financial support during his visit at Universidade
de S\~ao Paulo.  We are grateful to Mike West, Bas\'{\i}lio Santiago,
Arunav Kundu and Keith Ashman, the paper referee, for the careful reading
of this manuscript and useful comments to improve its content and to
H\'ector Cuevas and Leopoldo Infante for observing the standard stars
for calibration of the NGC 6868 field.

\newpage

% Figure Captions

% Figure 1
\figcaption[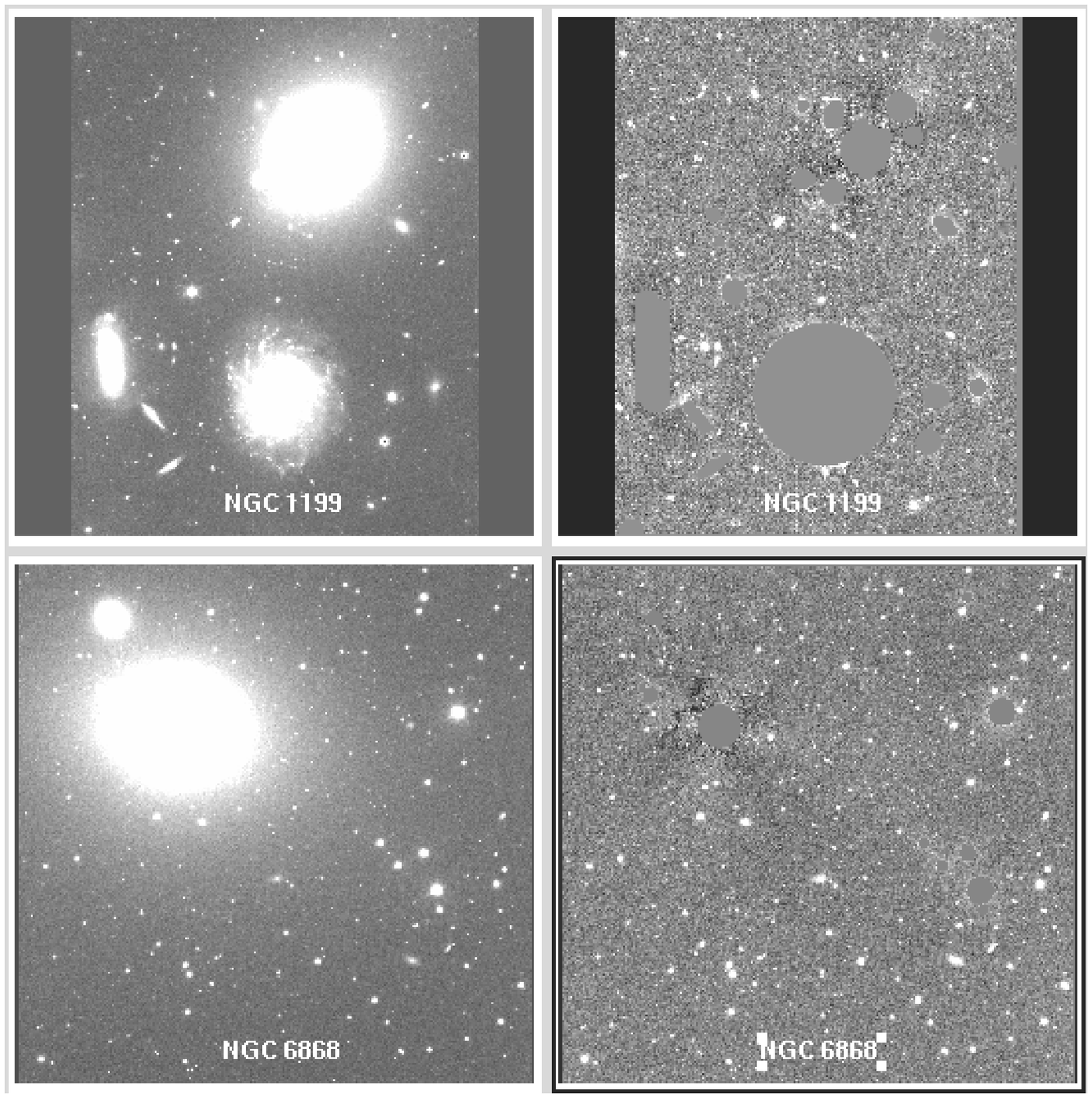]{The left panels show the combined images
for HCG 22 (upper panel) and NGC 6868 (lower panel) in the $R$ band. The
right panels show the final background subtracted masked images for
HCG 22 (upper panel) and NGC 6868 (lower panel) where the detection and
photometry were performed. The field sizes for the upper panels are $5.7
\times 7.3$ arcmin and $6.4 \times 6.4$ arcmin for the lower panels.
\label{figima}}

% Figure 2
\figcaption[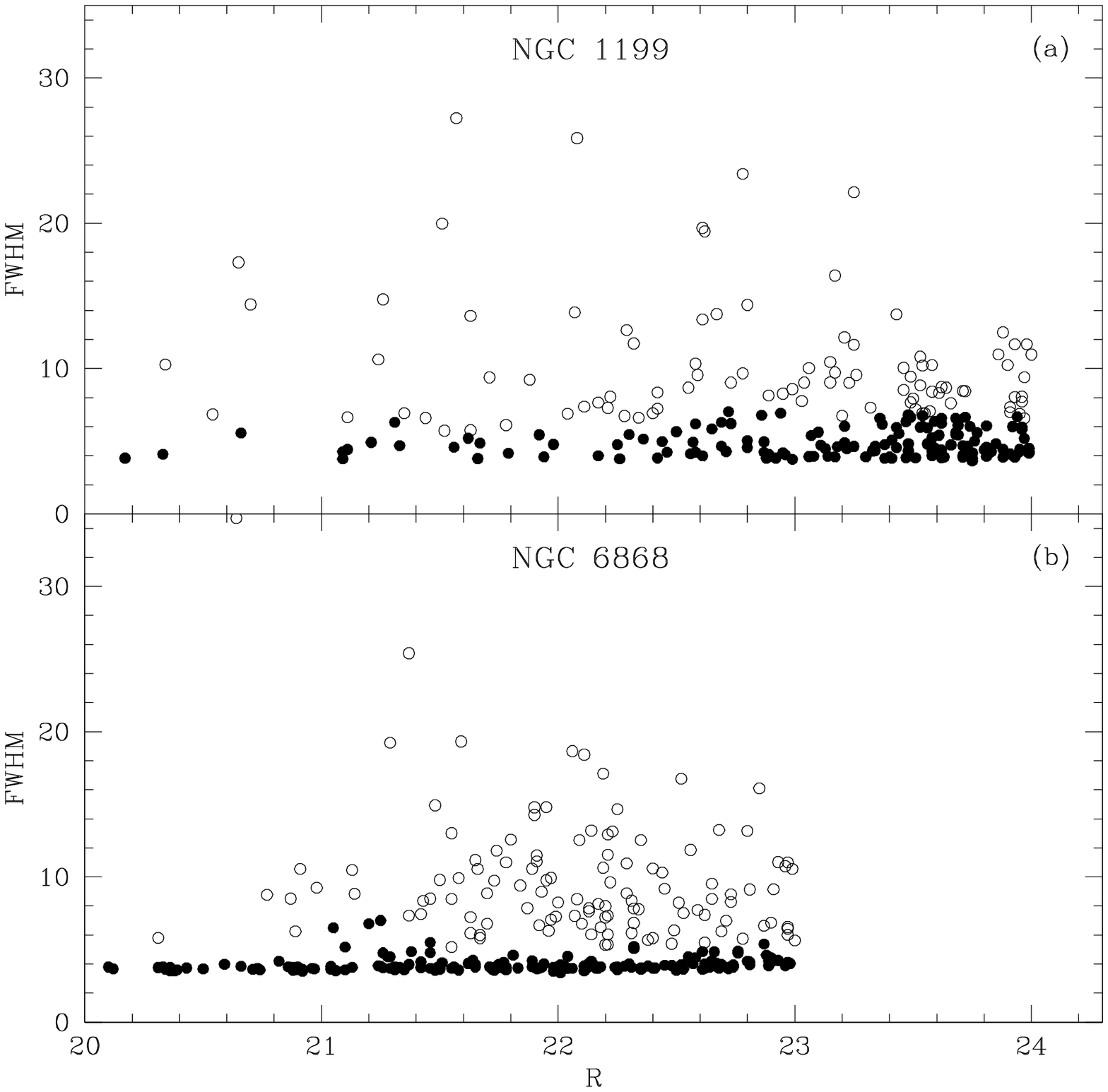]{Magnitude--FWHM plane used to select
point--like sources. The filled circles are the objects classified as
point--like sources and the open circles are the objects classified as
extended objects. Panel (a) shows the objects detected in the NGC 1199
images. Panel (b) shows the objects detected in the NGC 6868 images.
\label{figfwhm}}

% Figure 3
\figcaption[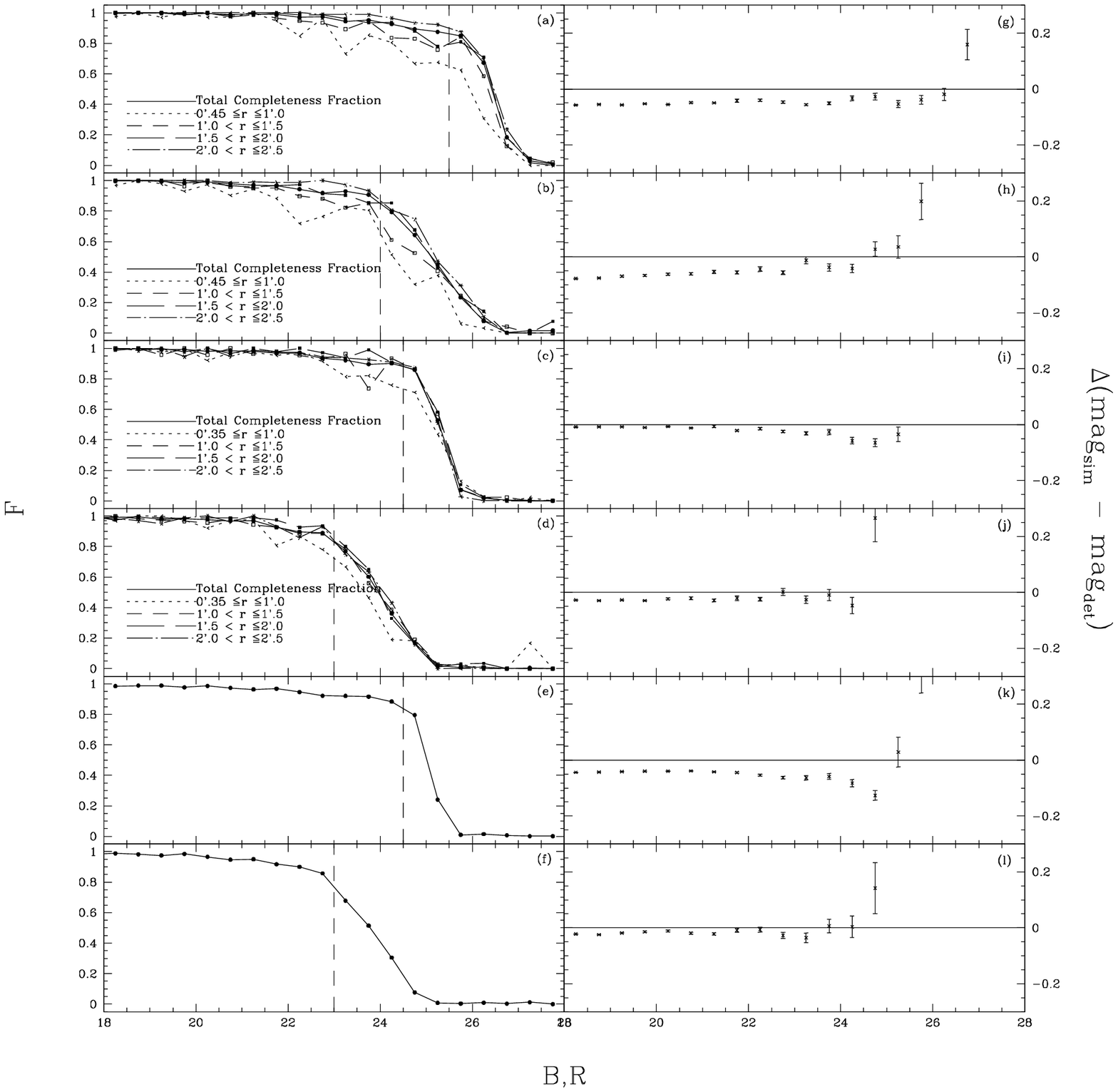]{The left panels show the detection
completeness fractions estimated with the add--star experiment performed
in the six images of our study. Panels (a) and (b) give the total
and radial completeness fractions for the HCG 22 images in the $B$
and $R$ bands, respectively. Panels (c) and (d) give the fractions
for the NGC 6868 images in the $B$ and $R$ bands and panels (e) and
(f) give the total completeness fraction for the NGC 6868 background
fields in the $B$ and $R$ bands, respectively. The right panels show
the photometric errors estimated with the add--star experiment. Panels
(g) and (h) give the errors for the HCG 22 images in the $B$ and $R$
bands, respectively.  Panels (i) and (j) for the NGC 6868 images and (k)
and (l) for the NGC 6868 background fields in the $B$ and $R$ bands,
respectively. The vertical dashed lines  in the left panels mark the
faint magnitude limit considered in each sample.  \label{figcompl}}

% Figure 4
\figcaption[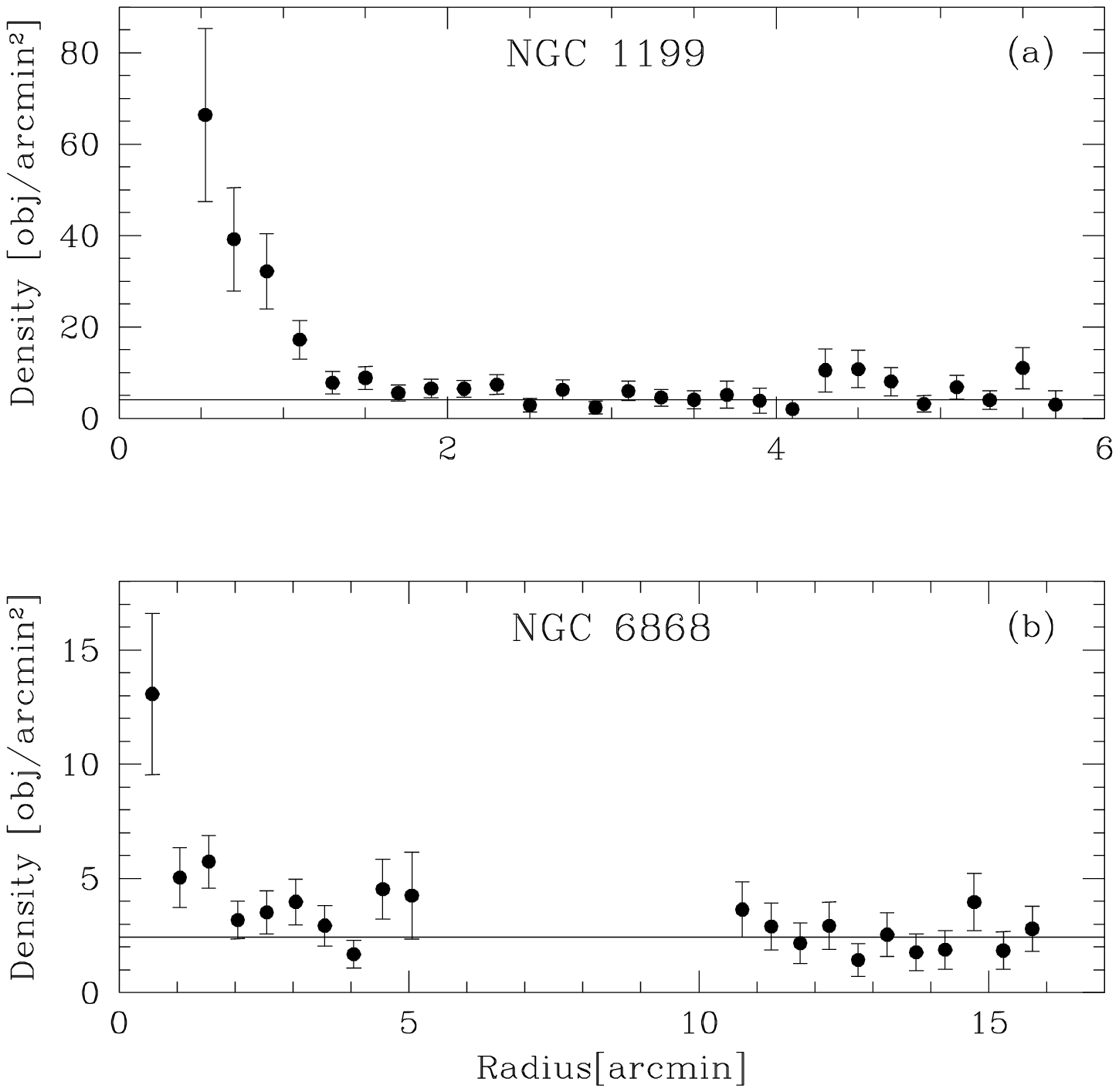]{Radial profile of objects classified
as point--like sources within the magnitudes and color limits ($0.7
\le~(B-R)_0 \le~2.1$). Panel (a) shows the distribution of objects around
NGC 1199, from 0 to 6 arcmin in rings of 0.2 arcmin. The magnitude ranges
are $B = 22.0$ to $25.5$ and $R = 20.0$ to $24.0$. Panel (b) shows the
distribution of objects around NGC 6868, from 0 to 17 arcmin in rings
of 0.5 arcmin. The magnitude ranges are $B = 22.0$ to $24.5$ and $R =
20.0$ to $23.0$. The solid lines represent the estimated background level:
$4.2~{\rm objects/arcmin^2}$ for NGC 1199 and $2.4~{\rm objects/arcmin^2}$
for NGC 6868. Note the different scales for the upper and lower panel.
\label{figrad}}

% Figure 5
\figcaption[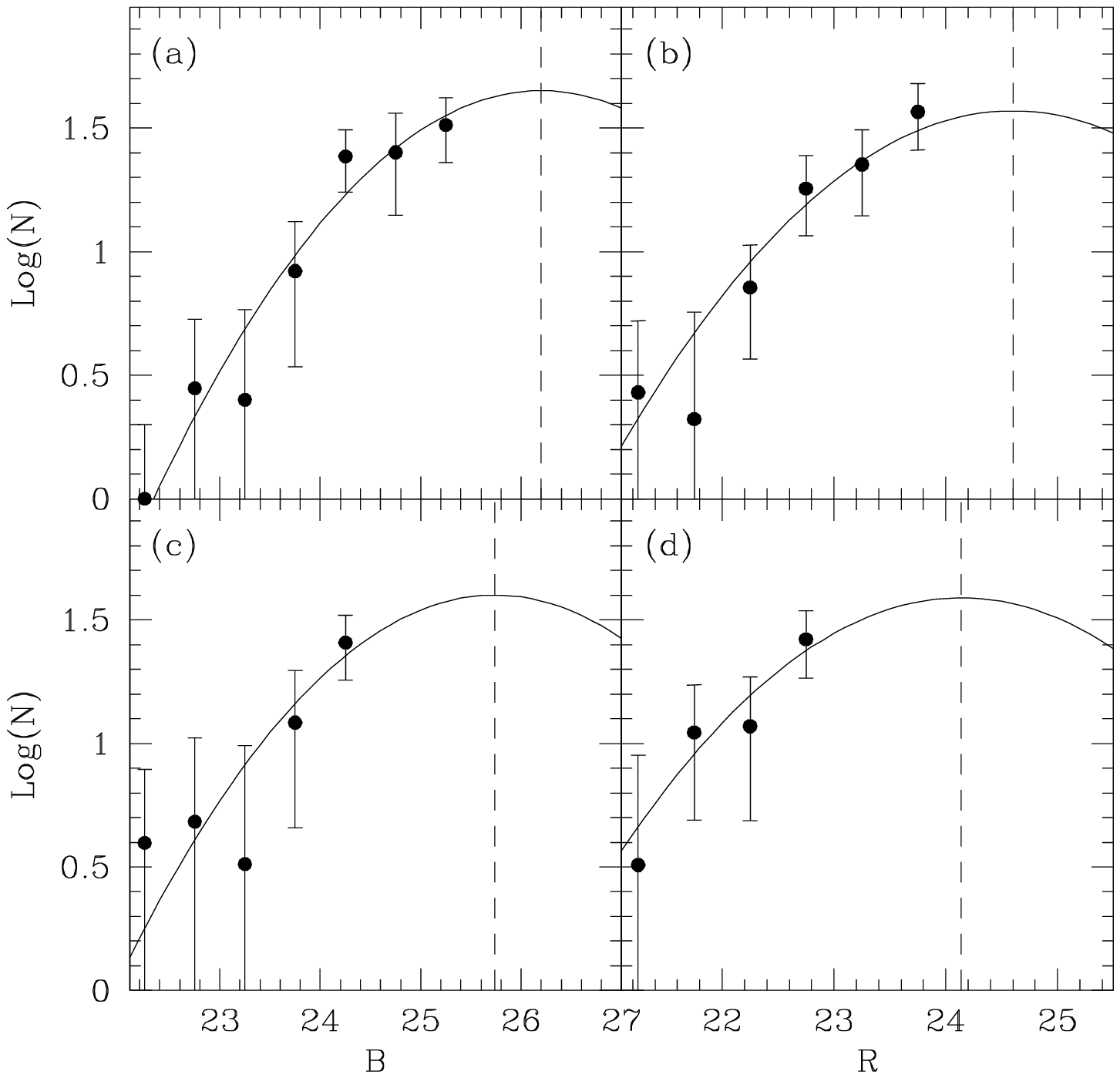]{Globular cluster luminosity functions. Panel
(a) shows the luminosity function of the globular cluster system around
NGC 1199 in the $B$ band and panel (b) in the $R$ band. Panels (c) and (d)
show the luminosity function of the globular cluster system around NGC
6868 in the $B$ and $R$ band respectively. The solid line represents de
Gaussian fit and the dashed lines are the expected turnover magnitudes.
\label{figlf}}

% Figure 6
\figcaption[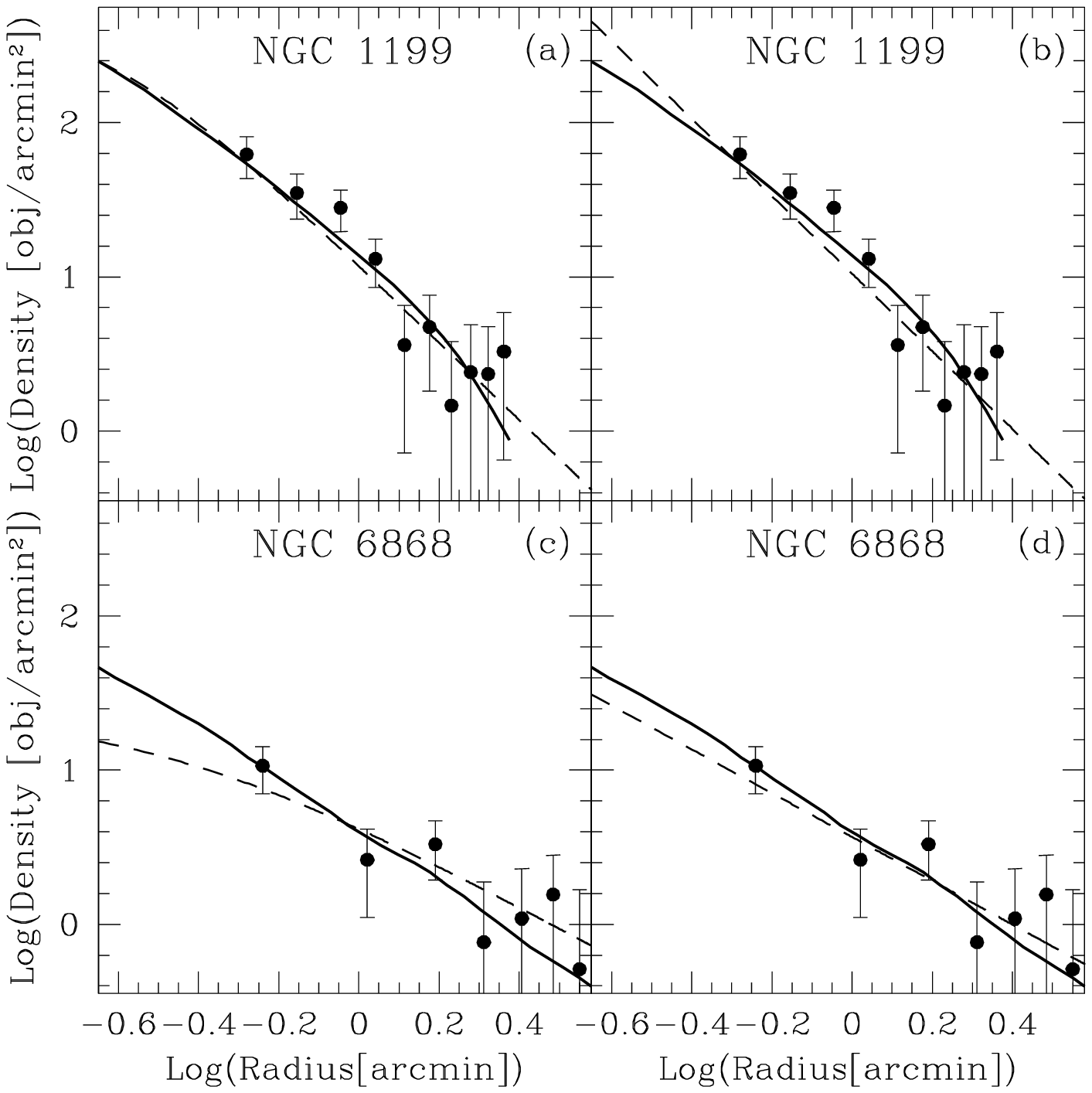]{Radial profiles of globular cluster
candidates. Panel (a) shows the radial profile of the candidates around
NGC 1199 with the fitted ``core model'' profile with the best--suited
core radius \citep{for96} (dashed line) and the galaxy light in the
B band arbitrarily vertically shifted (continuous line) overplotted.
Panel (b) shows the radial profile of the GCS around NGC 1199 and the
galaxy light (continuous line) with the power--law profile (dashed
line) overplotted. Panels (c) and (d) show the same information of
panels (a) and (b), respectively, for the candidates around NGC 6868.
\label{figking}}

% Figure 7
\figcaption[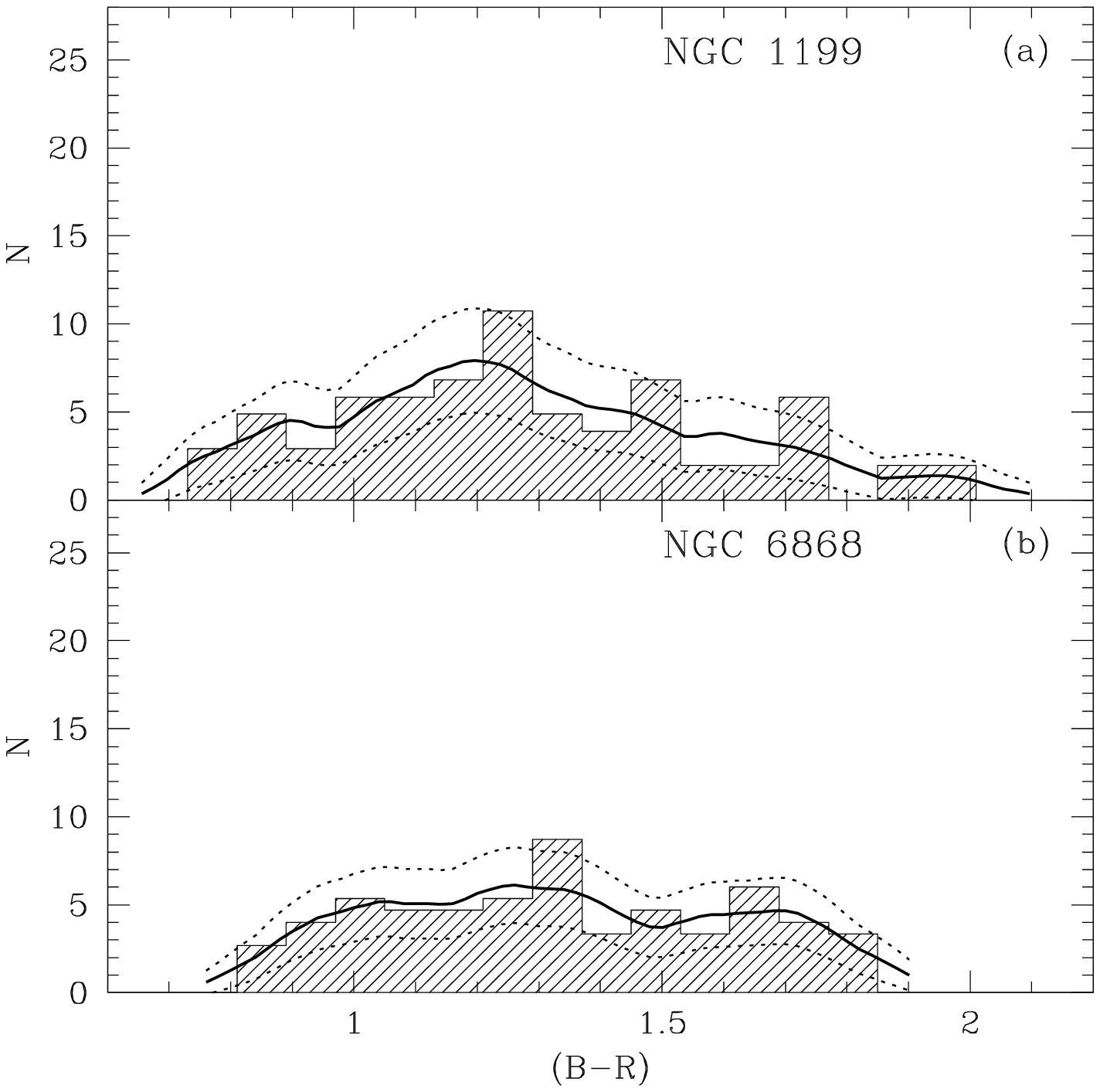]{Color distribution of background objects.
Panel (a) shows the color distribution of objects in the background
area of NGC 1199 and panel (b) of those in the background area of NGC
6868. The solid line represents the distribution function derived by
the Epanechnikov kernel density estimator and the dotted lines are its
upper and lower limits.  \label{figcolorbkg}}

% Figure 8
\figcaption[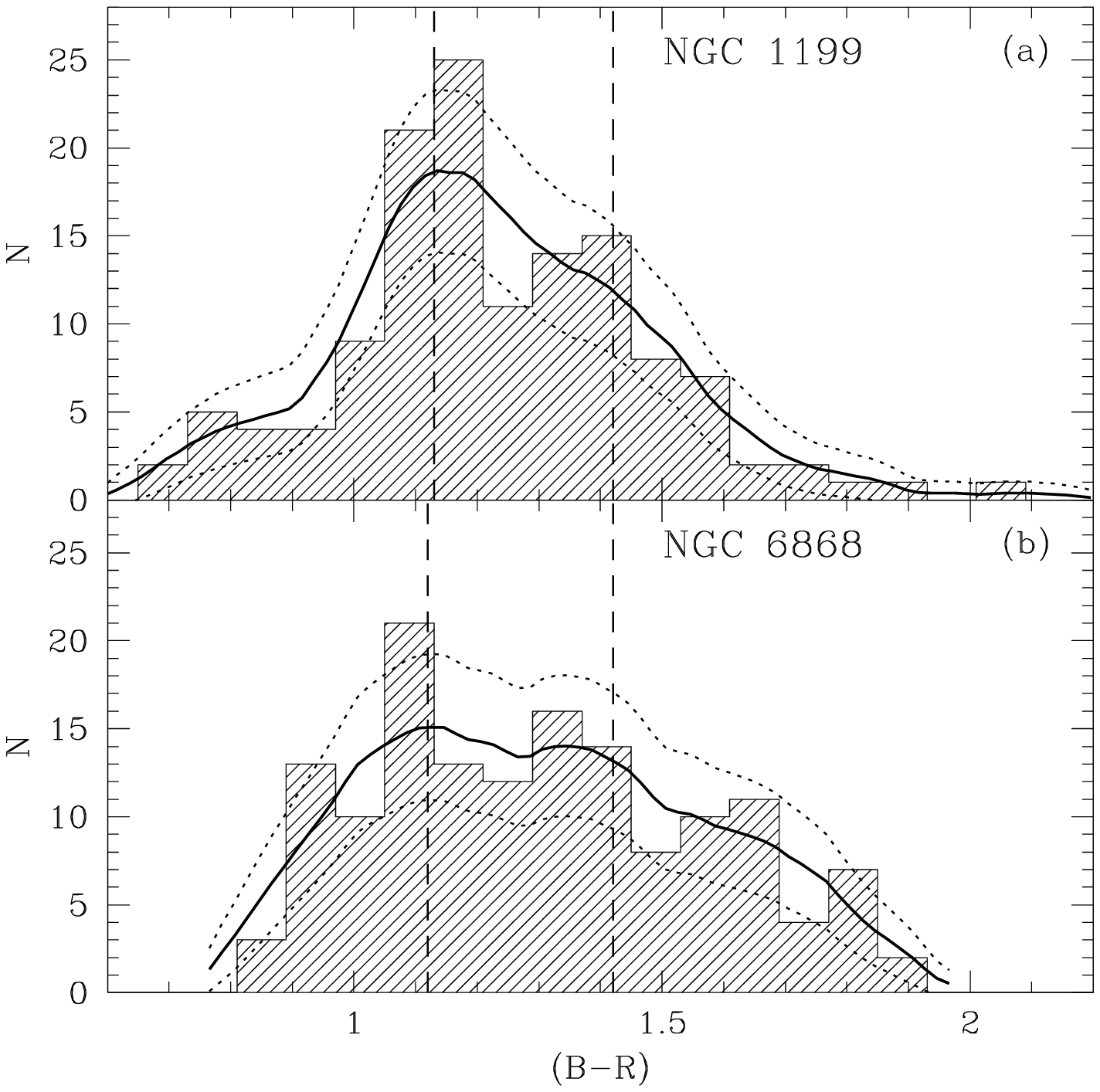]{Color distribution of globular cluster
candidates (no background subtraction is done). Panel (a) shows the color
distribution of objects around NGC 1199 and panel (b) of those around
NGC 6868. The solid line represents the distribution function derived
by the Epanechnikov kernel density estimator and the dotted lines are
its upper and lower limits. The vertical dashed lines represent the peak
values found by KMM.  \label{figcolor}}

% Figure 9
\figcaption[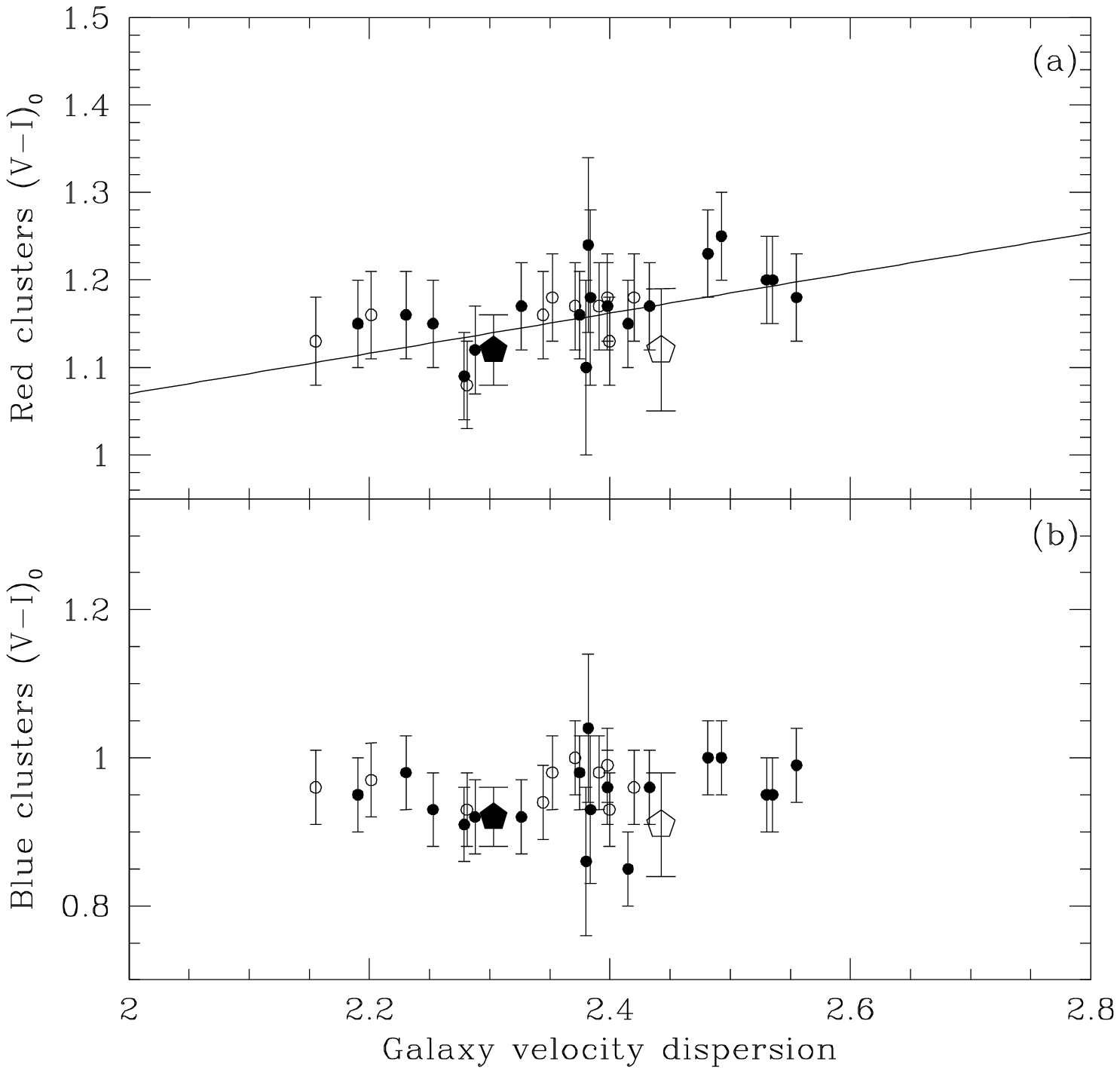]{Color--velocity dispersion relation for the
early--type galaxies of \citet{for01}. Panel (a) shows the values for the
red peaks and panel (b) for the blue peaks. The solid line represents
the color--velocity dispersion relation proposed by \citet{for01}.
The filled pentagons are the values for NGC 1199 and the open ones are
the values for NGC 6868.  \label{figforbes}}

\clearpage

% Figures

% Figure 1
\begin{figure}
\psfig{file=DaRocha.fig1.ps,angle=0,width=18.0 cm} 
\centerline{Figure 1.}
\end{figure}

\clearpage

% Figure 2
\begin{figure}
\psfig{file=DaRocha.fig2.ps,angle=0,width=12.0 cm} 
\centerline{Figure 2.}
\end{figure}

\clearpage

% Figure 3
\begin{figure} 
\psfig{file=DaRocha.fig3.ps,angle=0,width=18.0 cm} 
\centerline{Figure 3.}
\end{figure}

\clearpage

% Figure 4
\begin{figure} 
\psfig{file=DaRocha.fig4.ps,angle=0,width=12.0 cm} 
\centerline{Figure 4.}
\end{figure}

\clearpage

% Figure 5
\begin{figure} 
\psfig{file=DaRocha.fig5.ps,angle=0,width=12.0 cm}
\centerline{Figure 5.}
\end{figure}

\clearpage

% Figure 6
\begin{figure} 
\psfig{file=DaRocha.fig6.ps,angle=0,width=12.0 cm} 
\centerline{Figure 6.}
\end{figure}

\clearpage

% Figure 7
\begin{figure}
\psfig{file=DaRocha.fig7.ps,angle=0,width=12.0 cm}
\centerline{Figure 7.}
\end{figure}

\clearpage

% Figure 8
\begin{figure} 
\psfig{file=DaRocha.fig8.ps,angle=0,width=12.0 cm} 
\centerline{Figure 8.}
\end{figure}

\clearpage

% Figure 9
\begin{figure}
\psfig{file=DaRocha.fig9.ps,angle=0,width=12.0 cm}
\centerline{Figure 9.}
\end{figure}

\clearpage

% Table 1
\begin{deluxetable}{lll}

\footnotesize
\tablecaption{General properties of the observed galaxies.
\label{tabdata}}
\tablecolumns{3}
\tablewidth{0pt}
\tablehead{
\colhead{} & \colhead{NGC 1199} & \colhead{NGC 6868} 
}

\startdata
l                   & 199.2                & 350.9                \\
b                   & -57.3                & -32.6                \\
RA                  & 03:03:38.6           & 20:09:54.1           \\
DEC                 & -15:36:51            & -48:22:47            \\
Morph. Type         & E2                   & E2                   \\
$B_{T}$ & 12.5$\pm$0.13\tablenotemark{a} & 11.4$\pm$0.11\tablenotemark{b} \\
$M_V$               & -21.3$\pm$0.34       & -21.9$\pm$0.29       \\
$(B-V)$ & 0.99\tablenotemark{b} & 1.00\tablenotemark{b} \\
$V_{Rad}$ & 2705 ($km~s^{-1}$)\tablenotemark{b} & 2876 ($km~s^{-1}$)\tablenotemark{b} \\
Distance & 33.1 ($Mpc$)\tablenotemark{c} & 26.8 ($Mpc$)\tablenotemark{c} \\
Effective Radius  & 33.6$^{"}$\tablenotemark{d} & 38.4$^{"}$\tablenotemark{e} \\
Int. Vel. Disp. & 201 ($km~s^{-1})$\tablenotemark{f} & 277 ($km~s^{-1}$)\tablenotemark{f} \\

\enddata

\tablenotetext{a}{\citet*{hic89}}
\tablenotetext{b}{\citet{fab89}}
\tablenotetext{c}{\citet{ton01}}
\tablenotetext{d}{\citet{zep93}}
\tablenotetext{e}{\citet{kob99}}
\tablenotetext{f}{\citet{pru96}}

\end{deluxetable}

\clearpage

% Table 2
\begin{deluxetable}{lclllclcc}
\footnotesize
\tablecaption{Observational Information \label{tabimag}}
\tablewidth{0pt}
\tablehead{
\colhead{Group} & \colhead{Band}   & \colhead{Telesc.}   &
\colhead{Instr.} & \colhead{Date} & \colhead{Tot. Exp.} &
\colhead{Pixel Size} & \colhead{\# Ima.} & \colhead{Seeing}\\
\colhead{} & \colhead{}   & \colhead{}   & 
\colhead{} & \colhead{} & \colhead{[sec]} &
\colhead{} & \colhead{} & \colhead{['']}
}

\startdata
HCG 22      &B  &Keck II &LRIS  &06/02/1997 &720 &0.215 &4 &0.77\\
HCG 22      &R  &Keck II &LRIS  &06/02/1997 &630 &0.215 &7 &0.74\\
NGC 6868 (1)&B  &VLT     &FORS1 &10/10/1999 &900 &0.2 &5 &0.76\\
NGC 6868 (1)&R  &VLT     &FORS1 &10/10/1999 &810 &0.2 &9 &0.73\\
NGC 6868 (2)&B  &VLT     &FORS1 &10/11/1999 &900 &0.2 &5 &0.96\\
NGC 6868 (2)&R  &VLT     &FORS1 &10/11/1999 &810 &0.2 &9 &0.89\\
\enddata

\end{deluxetable}

\clearpage

% Table 3
\begin{deluxetable}{lccc}

\footnotesize
\tablecaption{Detection completeness limit with the percentage of 
classification. Column (1) Image name, column (2) image band, 
column (3) completeness limit magnitude, column (4) percentage
of detection at the completeness limit magnitude.
\label{tablim}}
\tablecolumns{4}
\tablewidth{0pt}
\tablehead{
\colhead{} & \colhead{} & \multicolumn{2}{c}{Detec. Completeness} \\
\cline{3-4} \\
\colhead{Image} & \colhead{band} & \colhead{Mag. Limit} & \colhead{\%} 
}

\startdata
HCG 22      & B & 25.5 & 87.5 \\
HCG 22      & R & 24.0 & 79.5 \\
NGC 6868 (1)& B & 24.5 & 90.2 \\
NGC 6868 (1)& R & 23.0 & 88.9 \\
NGC 6868 (2)& B & 24.5 & 88.3 \\
NGC 6868 (2)& R & 23.0 & 85.7 \\
\enddata

\end{deluxetable}

\clearpage

% Table 4
\begin{deluxetable}{lcrrr}

\footnotesize
\tablecaption{Gaussian fit parameters. Column (1) image name, column
(2) image band, column (3) peak position (fixed at $M_V=-7.33$), column
(4) $\chi^2$ of the Gaussian fit, column (5) area of Gaussian covered
by our data.
\label{tabgauss}}
\tablecolumns{6}
\tablewidth{0pt}
\tablehead{
\colhead{Image} & \colhead{Band} & \colhead{$m_0$} & \colhead{$\chi^2$} & \colhead{\%}
}

\startdata
HCG 22   & B & $26.2\pm0.3$ & 0.3214 & $30.8\pm8.5$ \\
HCG 22   & R & $24.6\pm0.3$ & 0.2630 & $33.4\pm8.7$ \\
NGC 6868 & B & $25.7\pm0.2$ & 0.2854 & $18.7\pm5.3$ \\
NGC 6868 & R & $24.1\pm0.2$ & 0.1995 & $20.9\pm5.2$ \\
\enddata

\end{deluxetable}

\clearpage

% Table 5
\begin{deluxetable}{lrr}

\footnotesize
\tablecaption{Specific frequency values using the ``core model'' profile
with the core radius ($r_c$) proposed by \citet{for96} and a power--law
profile.  Column (1) estimative (Local values, core radius value in $kpc$
for the ``core model'' profile and slope of the power--law profile,
column (2) estimated number of globular clusters, column (3) estimated
specific frequency.
\label{tabsn}}
\tablecolumns{3}
\tablewidth{0pt}
\tablehead{
\colhead{} & \colhead{$N$} & \colhead{$S_n$} 
}

\startdata
\cutinhead{NGC 1199}
Local                &   $314\pm105$ & $3.4\pm1.5$ \\
$r_c = 2.19$         &  $1169\pm429$ & $3.6\pm1.8$ \\
$\alpha = 2.5\pm0.3$ &  $1668\pm899$ & $5.2\pm3.2$ \\
\cutinhead{NGC 6868}
Local                &   $266\pm106$ & $0.8\pm0.4$ \\
$r_c = 2.58$         &  $1060\pm555$ & $1.8\pm1.1$ \\
$\alpha = 1.4\pm0.3$ &  $1089\pm503$ & $1.9\pm1.0$ \\
\enddata

\end{deluxetable}

\end{document}